\documentclass[pra,superscriptaddress,twocolumn,floatfix,showpacs,nofootinbib]{revtex4-1}
\usepackage{graphicx,color}
\usepackage{amsmath,amssymb,bm}
\usepackage{braket}		% Dirac notation

\newcommand{\tn}{\textnormal}
\newcommand{\cprb}[3]{Phys.~Rev.~B {\bf #1}, #2 (#3)}

\newcommand{\fref}[1]{Fig.~\ref{#1}}
\newcommand{\eref}[1]{Eq.~\eqref{#1}}

\newcommand{\sref}[1]{Sec.~\ref{#1}}

\newcommand{\avg}[1]{\left\langle{#1}\right\rangle}
\newcommand{\C}{{\mathcal{C}}}
\newcommand{\dc}{\delta_\C}
\newcommand{\intc}[1]{\int_\C\mathrm{d}{#1}\,}
\newcommand{\chat}[1]{\ensuremath{\xop{#1}}}
\newcommand{\tempop}[3][\textstyle]{\settowidth{\dimen1}{$#1\hat{#2}$}\makebox[\dimen1][l]{$#1\hat{#2\mspace{#3}}$}}
\newcommand{\xop}[1]{{\mathchoice{\tempop[\displaystyle]{#1}{3.5mu}}{\tempop{#1}{3.5mu}}{\tempop[\scriptstyle]{#1}{3.5mu}}{\tempop[\scriptscriptstyle]{#1}{3mu}}}}
\newcommand{\cbar}[1]{\ensuremath{\xbar{#1}}}
\newcommand{\cbarbar}[1]{\ensuremath{\xbarbar{#1}}}
\newcommand{\tempbarbar}[3][\textstyle]{\settowidth{\dimen1}{$#1\bar{\bar{#2}}$}\makebox[\dimen1][l]{$#1\bar{\bar{#2\mspace{#3}}}$}}
\newcommand{\tempbar}[3][\textstyle]{\settowidth{\dimen1}{$#1\bar{#2}$}\makebox[\dimen1][l]{$#1\bar{#2\mspace{#3}}$}}
\newcommand{\xbarbar}[1]{{\mathchoice{\tempbarbar[\displaystyle]{#1}{3.5mu}}{\tempbarbar{#1}{3.5mu}}{\tempbarbar[\scriptstyle]{#1}{3.5mu}}{\tempbarbar[\scriptscriptstyle]{#1}{3mu}}}} 
\newcommand{\xbar}[1]{{\mathchoice{\tempbar[\displaystyle]{#1}{3.5mu}}{\tempbar{#1}{3.5mu}}{\tempbar[\scriptstyle]{#1}{3.5mu}}{\tempbar[\scriptscriptstyle]{#1}{3mu}}}} 
\renewcommand{\d}{{\mathrm{d}}}
\renewcommand{\i}{{\mathrm{i}}}
\newcommand{\lowerbossphantom}{\vphantom{\cbarbar{x}}}
\newcommand{\upperbossphantom}{\vphantom{\dagger}}
\newcommand{\aop}[2]{\ensuremath{\chat{c}_{#1,#2\lowerbossphantom}^{\upperbossphantom}}}
\newcommand{\cop}[2]{\ensuremath{\chat{c}_{#1,#2\lowerbossphantom}^{\dagger\upperbossphantom}}}
\newcommand{\G}[3]{\ensuremath{G_{#1#2\lowerbossphantom}^{#3\upperbossphantom}}}

\newcommand{\n}[2]{\ensuremath{n_{#1\lowerbossphantom}^{#2\upperbossphantom}}}
\newcommand{\spdm}[3]{\ensuremath{n_{#1,#2\lowerbossphantom}^{#3\upperbossphantom}}}
\newcommand{\doc}[2]{\ensuremath{d_{#1\lowerbossphantom}^{#2\upperbossphantom}}}

\newcommand{\T}[2]{\ensuremath{T_{#1#2\lowerbossphantom}^{\upperbossphantom}}}
\renewcommand{\S}[3]{\ensuremath{\Sigma_{#1#2\lowerbossphantom}^{#3\upperbossphantom}}}
\newcommand{\h}[3]{\ensuremath{h_{#1#2\lowerbossphantom}^{#3\upperbossphantom}}}
\newcommand{\kronecker}[2]{\delta^{\upperbossphantom}_{{#1},{#2}\lowerbossphantom}}
\newcommand{\sx}{s}
\newcommand{\sxbm}{\bm{\sx}}
\newcommand{\nsx}{N_{\mathrm{s}}}

\newcommand{\spbm}{{\sxbm}^{\prime}}
\renewcommand{\sb}{\cbar{s}}
\newcommand{\sbbm}{\bm{\sb}}

\newcommand{\Sx}{\sigma}

\begin{document}
\title{Nonequilibrium dynamics in the one-dimensional Fermi-Hubbard model: A comparison of the nonequilibrium Green functions approach
and the density matrix renormalization group method}

\author{N. Schl\"unzen}
\affiliation{Institut f\"ur Theoretische Physik und  Astrophysik, Christian-Albrechts Universit\"at Kiel, Leibnizstr. 15, Germany}
\author{J.-P. Joost}
\affiliation{Institut f\"ur Theoretische Physik und  Astrophysik, Christian-Albrechts Universit\"at Kiel, Leibnizstr. 15, Germany}
\author{F. Heidrich-Meisner}
\affiliation{Department of Physics and Arnold Sommerfeld Center for Theoretical Physics,Ludwig-Maximilians-Universit\"at M\"unchen, D-80333 M\"unchen, Germany}
\author{M. Bonitz}
\affiliation{Institut f\"ur Theoretische Physik und  Astrophysik, Christian-Albrechts Universit\"at Kiel, Leibnizstr. 15, Germany}

\begin{abstract}
The nonequilibrium dynamics of strongly-correlated fermions in lattice systems have attracted considerable interest in the condensed matter and ultracold atomic-gas communities. 
While experiments have made remarkable progress in recent years, there remains a need for the further development of theoretical tools that can account for
both the nonequilibrium conditions and strong correlations. For instance, 
time-dependent theoretical quantum approaches based on the density matrix renormalization group (DMRG) methods  have been primarily 
applied to one-dimensional setups. Recently, two-dimensional quantum simulations of the expansion of fermions based on nonequilibrium Green functions (NEGF) have been presented [Schl\"unzen {\em et al.}, \cprb{93}{035107}{2016}] that showed excellent agreement with the experiments. Here we present an extensive comparison of the NEGF approach to  numerically accurate DMRG results. The results indicate that  NEGF are a reliable theoretical tool for weak to intermediate coupling strengths in arbitrary dimensions and make long simulations possible. This is complementary to DMRG simulations which are particularly efficient at strong coupling.
\end{abstract}

% 67.85.-d 	Ultracold gases, trapped gases
% 05.30.Jp 	Boson systems
% 37.10.Jk 	Atoms in optical lattices
%%%%%%%%%%%%%%%%%%%%%%%%%%%%%%%%%%%%%%%%%%%%%%%%%%%%%%%%%%%%%%%%%%%%%%%%%%%%%%%

\maketitle

\section{Introduction}\label{s:intro}

Experiments addressing the nonequilibrium dynamics of quantum many-body systems have made remarkable progress in recent years, both probing ultrafast dynamics in strongly correlated materials \cite{Giannetti2016,Gandolfi2016} and
quantum quenches in interacting quantum gases (see Refs.~\onlinecite{langen15,eisert15,gogolin15} for a review). 
Among the many  ultracold quantum-gas experiments with fermions we mention 
the study of the 
expansion dynamics of strongly-correlated fermions in a two-dimensional optical lattice 
 \cite{schneider12}, the collapse and revival dynamics of Fermi-Bose mixtures \cite{will14} and the real-time decay of
a density wave in one-dimensional lattices \cite{pertot14}.
Very recently, several experimental groups reported the successful implementation of fermionic quantum-gas microscopes \cite{Cheuk2015,Parsons2015,haller2015,edge2015,Omran2015,Cheuk2016,Cheuk2016a,Greif2016,boll2016}, which will give unprecedented access to both equilibrium and nonequilibrium properties of the Fermi-Hubbard model.
Given the tremendous success of the earlier bosonic quantum-gas microscopes in exploring the nonequilibrium realm \cite{cheneau12,fukuhara13,fukuhara13a,hild14,preiss15}, a considerable experimental activity in studying quantum-quench dynamics in the Fermi-Hubbard model can be expected in the near future.
Quantum-gas microscopes operate with two-dimensional systems which will push the efforts into this most challenging regime (see also Ref.~\onlinecite{Cocchi2016}) while also
allowing to study one-dimensional systems \cite{boll2016}.

A large body of theoretical work has concentrated on one-dimensional systems, the reason being both experiments \cite{kinoshita06,Hofferberth2007,Langen2013,Gring2012,trotzky12,cheneau12,ronzheimer13} as well as the 
availability of powerful theoretical tools based on field theory \cite{giamarchi}, integrability \cite{essler-book} or numerical methods. 
While exact diagonalization (ED) is still an indispensable tool (see, e.g., Refs.~\onlinecite{manmana05,rigol08,prelovsek11}), it is limited to small systems. 
Nonetheless, for problems restricted to the dynamics of a single charge carrier coupled to spin or phonon degrees of freedom, there exist Krylov-space approaches that 
operate in a subspace of the full Hilbert space constructed by  selecting only those states accessible
by the Hamiltonian dynamics \cite{bonca99}. Such an exact diagonalization in a limited functional space has been applied quite extensively
to two-dimensional problems as well (see, e.g., Refs.~\onlinecite{vidmar11,mierzejewski11,bonca12,golez14}).

Time-dependent density matrix renormalization group (DMRG) methods \cite{Vidal04,daley04,white04} have been very widely applied 
to nonequilibrium problems and yield numerically accurate results, but are limited by the accessible times scales and are 
primarily useful for one-dimensional systems. A recent variant of the method \cite{zaletel14} 
has been tailored for long-range interactions and is thus better 
suited for coupled one-dimensional and two-dimensional systems  \cite{zaletel14,Hauschild2015} but cannot overcome the exponential scaling of a matrix-product states
ansatz with the number of coupled chains.
The application of time-dependent tensor-network approaches that are based on ansatz states suitable for two-dimensional systems
such as the projected entangled pair states has been very little explored \cite{Dorando2009,Haegeman2011,lubasch11}.

Apart from time-dependent DMRG methods, there are other many-body methods for the real-time evolution including 
continuous-time quantum Monte Carlo \cite{gull11} and time-dependent dynamical mean-field theory approaches \cite{schmidt06,freericks06,eckstein10}. The former, while being able to achieve essentially
exact results for short evolution times, can suffer from a dynamical sign problem \cite{gull11}.
The latter method often utilizes
continuous-time quantum Monte Carlo as an impurity solver, while in more recent developments, time-dependent DMRG has also been successfully used for this purpose \cite{wolf14,ganahl15}. Time-dependent DMFT methods are not exact in two dimensions either but are argued to capture better the
physics of strongly-correlated systems in higher dimensions, leading to a wide range of applications in the context of 
nonequilibrium dynamics in the Hubbard model (see, e.g., Ref.~\onlinecite{eckstein10}). 
Finally, the iterative equation-of-motion method for operators provides an alternative approach \cite{uhrig09},
which has also been applied to quantum quench problems in the 2D Fermi-Hubbard model \cite{hamerla13}.

Despite all these efforts,  there still is  
a significant gap between the rapidly progressing experiments in the field of ultracold atoms and accurate quantum dynamics simulations when it comes to correlated systems in
two dimensions.
To contribute towards closing this gap, two of us have applied an alternative approach to the quantum simulation of the nonequilibrium mass-transport of correlated fermions studied in the experiment of Ref.~\onlinecite{schneider12}: nonequilibrium Green functions (NEGF). Previously, this theory has been successfully applied  to a variety of many-particle systems including the correlated electron gas \cite{kwong_prl_00},  electron-hole plasmas \cite{kwong_pss_98}, nuclear matter \cite{rios_11} and electrons in quantum dots \cite{balzer_prb_09, lorke_06}, for a recent overview see Ref.~\onlinecite{pngf6}.
Extensive applications to finite Hubbard clusters were presented in Ref.~\onlinecite{hermanns_prb_14} and first applications of NEGF to mass transport in small lattice systems of correlated fermions were shown in Ref.~\onlinecite{bonitz_cpp_15}. Finally, in Ref.~\onlinecite{schluenzen16} these simulations were extended to strong coupling by using $T$-matrix selfenergies as well as to substantially larger systems. Applying an extrapolation to the thermodynamic limit the nonequilibrium correlated quantum mass transport in two-dimensional fermion ensembles could be directly compared to the experiments of Ref.~\onlinecite{schneider12}, and excellent agreement was observed. For an overview on the NEGF approach and its application to inhomogeneous Hubbard clusters, see Ref.~\onlinecite{schluenzen_cpp_16}. 

Even though NEGF simulations are computationally demanding, they have a number of remarkable advantages. First, they do not exhibit an exponential scaling with system size, as is the case for exact diagonalization,  and they do not have a dynamical sign problem as continuous time QMC methods. Second, they are not limited with respect to the system dimensionality, as opposed to matrix-product state methods.
At the same time, in contrast to ED, NEGF simulations are not a first-principle method since they involve a many-body approximation---the selfenergy---which determines the accuracy and the quality of the results, similar to the approximate exchange-correlation energy in density functional theory. DMRG, on the other hand, also involves an approximation but the numerical errors  depend on a control parameter, the discarded weight, and whenever
this can be made sufficiently small, the results can become essentially exact for system sizes larger than what is accessible to ED \cite{schollwoeck05,schollwoeck11}. 

The accuracy of NEGF simulations of spatially inhomogeneous fermion systems was tested before for few-electron atoms \cite{balzer_pra_10} and  small Hubbard clusters \cite{friesen10} where exact diagonalization results are available. This analysis was extended to larger Hubbard systems, on the order of 10 sites in Refs.~\onlinecite{hermanns_prb_14, lacroix_prb_14}, revealing a high accuracy of simulations with second-order Born selfenergies, for weak coupling and moderate times (on the order of 20 inverse hopping times). However, the quality of the results for larger systems has remained open until now, due to the lack of reliable benchmark data. On the other hand, for small Hubbard clusters, also problems were reported: in the case of a strong excitation, two-time NEGF simulations were found to exhibit an unphysical damping of the dynamics \cite{friesen09, friesen10}. The origin of this behavior has been traced back to the selfconsistent nature of the used approximations. These deficiencies could be removed to a large extent by making the transition to single-time dynamics with the help of the generalized Kadanoff--Baym ansatz \cite{lipavski} with Hartree-Fock propagators (HF-GKBA) \cite{hermanns_prb_14}. 
\begin{figure}[tb]
\includegraphics[width=\columnwidth]{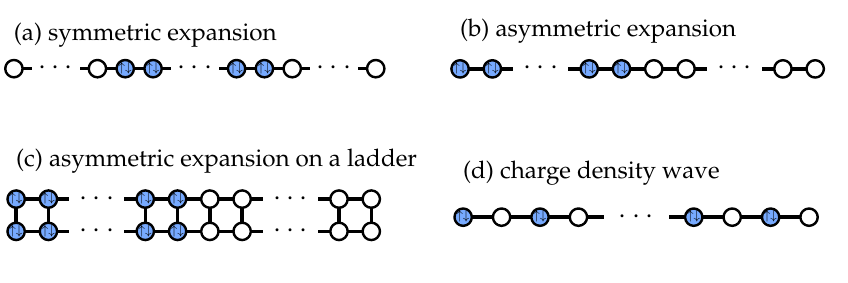}
\caption{
Initial states of the nonequilibrium problems studied in this paper: (a) Symmetric 1D sudden expansion from a band insulator (BI).
(b) Asymmetric 1D expansion from a BI. (c) Sudden expansion on a two-leg ladder. (d) Relaxation dynamics from a charge-density wave state 
$|\psi_0\rangle = | 2,0,2,0,2,0, \dots \rangle$ in 1D. The open circles indicate empty sites, the filled circles represent an initial occupation with 
two fermions, i.e., a doublon.
}\label{fig:initial_states}
\end{figure}

Thus, there is a clear need to further study the question of accuracy and predictive capability of NEGF simulations, in particular, for systems larger than those studied so far, for longer simulation times and beyond the weak-coupling limit.
The goal of this paper is to present such an analysis by benchmarking NEGF results using a  variety of different  selfenergy approximations, in a two-time as well as in a single-time formalism (i.e., using the GKBA), against  DMRG results. Due to the inherent properties of matrix-product states \cite{schollwoeck11},  
these comparisons have to focus on 1D fermion systems. We choose a set of four non-trivial cases of nonequilibrium dynamics in the Fermi-Hubbard model for which correlations play a crucial role. 

The Hamiltonian of the Fermi-Hubbard chain is
\begin{equation}
H = -J \sum_{\avg{\sxbm,\spbm}}\sum_{\Sx=\uparrow,\downarrow} \left(\cop{\sxbm}{\Sx}\aop{\spbm}{\Sx} +\tn{h.c.}\right) + U \sum_{\sxbm} \n{\sxbm}{\uparrow} \n{\sxbm}{\downarrow} \,,
\end{equation}
where $\cop{\sxbm}{\Sx}$ creates a fermion with spin $\Sx=\uparrow,\downarrow$ in site $\sxbm$ and $\n{\sxbm}{\Sx}= 
\cop{\sxbm}{\Sx}\aop{\sxbm}{\Sx}$. $J$ is the hopping matrix element (set to unity in our simulations), $U$ denotes the
onsite interaction and $L$ is the number of sites (the lattice spacing and $\hbar$ are set to unity).
The cases studied include (i) the symmetric and (ii) asymmetric expansion from a band insulator into an empty lattice, 
(iii) the expansion from a band insulator in a quasi-1D situation on  a two-leg ladder, and 
(iv) the decay of an ideal charge-density wave state.
 These four initial situations are sketched in  \fref{fig:initial_states}. 
 
 As a result of this analysis, the applicability range of NEGF simulations and relevant approximations is being mapped out. Our main results are the following: NEGF simulations with the HF-GKBA are reliable also for moderate coupling, $U/J \lesssim 4$, if the proper selfenergies are being used. These are the $T$-matrix selfenergy--for small or large filling--and the third-order selfenergy (including all diagrams of third order, cf. Sec.~\ref{sec:negf})--close to half filling. In all cases, two-time simulations are less accurate (due to the unphysical damping mentioned above) but they can be used to estimate the deviations of the HF-GKBA from the exact result as typically the latter is enclosed between single-time and two-time NEGF results. Finally, NEGF simulations fill the gap left open by DMRG by being capable to treat large systems (of any dimensionality) and to achieve long simulation times, for weak and moderate coupling, whereas the DMRG is advantageous and more efficient for strong coupling.
 
 %This will allow one to use, in the future, the complementarity of the two approaches. In particular, NEGF will be useful to access longer simulation times and 2D and 3D systems.

The remainder of this paper is as follows: in \sref{sec:methods} we give a brief introduction into NEGF and time-dependent DMRG simulations. This is followed in \sref{s:results} by numerical results. There we study the four cases introduced above and depicted in~\fref{fig:initial_states}: a symmetric and asymmetric sudden expansion (confinement quench) in 1D, Secs.~\ref{sec:exp_1D} and \ref{sec:asym}, respectively, the sudden expansion in a two-leg ladder, \sref{sec:ladder}, and a charge-density-wave initial state, \sref{sec:cdw}.

\section{Methods}
\label{sec:methods}
\subsection{Nonequilibrium Green functions (NEGF)}
\label{sec:negf}
The central quantity in the nonequilibrium Green functions theory is the (single-particle) Green function $G$. It is defined on the Schwinger--Keldysh contour~\cite{schwinger61,keldysh65} $\C$ via the time-ordering operator $T_\C$,
\begin{align}\label{eq:g-def}
 \G{\sxbm}{\spbm}{\Sx} (z,z') = -\frac{\i}{\hbar}\avg{T_\C\aop{\sxbm}{\Sx}(z)\cop{\spbm}{\Sx}(z')}\,,
\end{align}
where, $\avg{\dots}$ denotes the ensemble average. The Green function can be understood as a generalization of the nonequilibrium single-particle density matrix, $\spdm{\sxbm}{\spbm}{\Sx}(t)$, onto the two-time plane. Therefore, $G$ provides easy access not only to the observables related to $\spdm{\sxbm}{\spbm}{\Sx}$ but, in addition, also to the spectral properties of the system. However, the full $N$-particle information is not directly available from $G$, although, for example, the pair correlation function can be reconstructed from $G$~\cite{kobusch_13}.

The equations of motion for the single-particle Green function are the Keldysh--Kadanoff--Baym equations~\cite{kadanoff_baym},
\begin{align}
\label{eq:kbe}
&\left(\i\hbar \frac{\partial}{\partial z}\kronecker{\sxbm}{\sbbm}-\h{\sxbm}{\sbbm}{\Sx}\right)\G{\sbbm}{\spbm}{\Sx}(z,z') \\
 &\quad= \dc(z-z')\kronecker{\sxbm}{\spbm}+\intc{\cbar{z}}\S{\sxbm}{\sbbm}{\Sx}(z,\cbar{z})\G{\sbbm}{\spbm}{\Sx}(\cbar{z},z')\,,\nonumber
\end{align}
together with the adjoint equation ($h$ denotes the matrix element of the single-particle Hamiltonian). $\Sigma$ denotes the selfenergy which is the only unknown of the theory, and with an exact $\Sigma$ the method would be exact. In practice, the selfenergy has to be approximated for which systematic many-body schemes (e.g. Feynman diagrams) exist that are applicable in equilibrium as well as in nonequilibrium, via the use of the time contour. 

\begin{figure}[tb]
\includegraphics[width=\columnwidth]{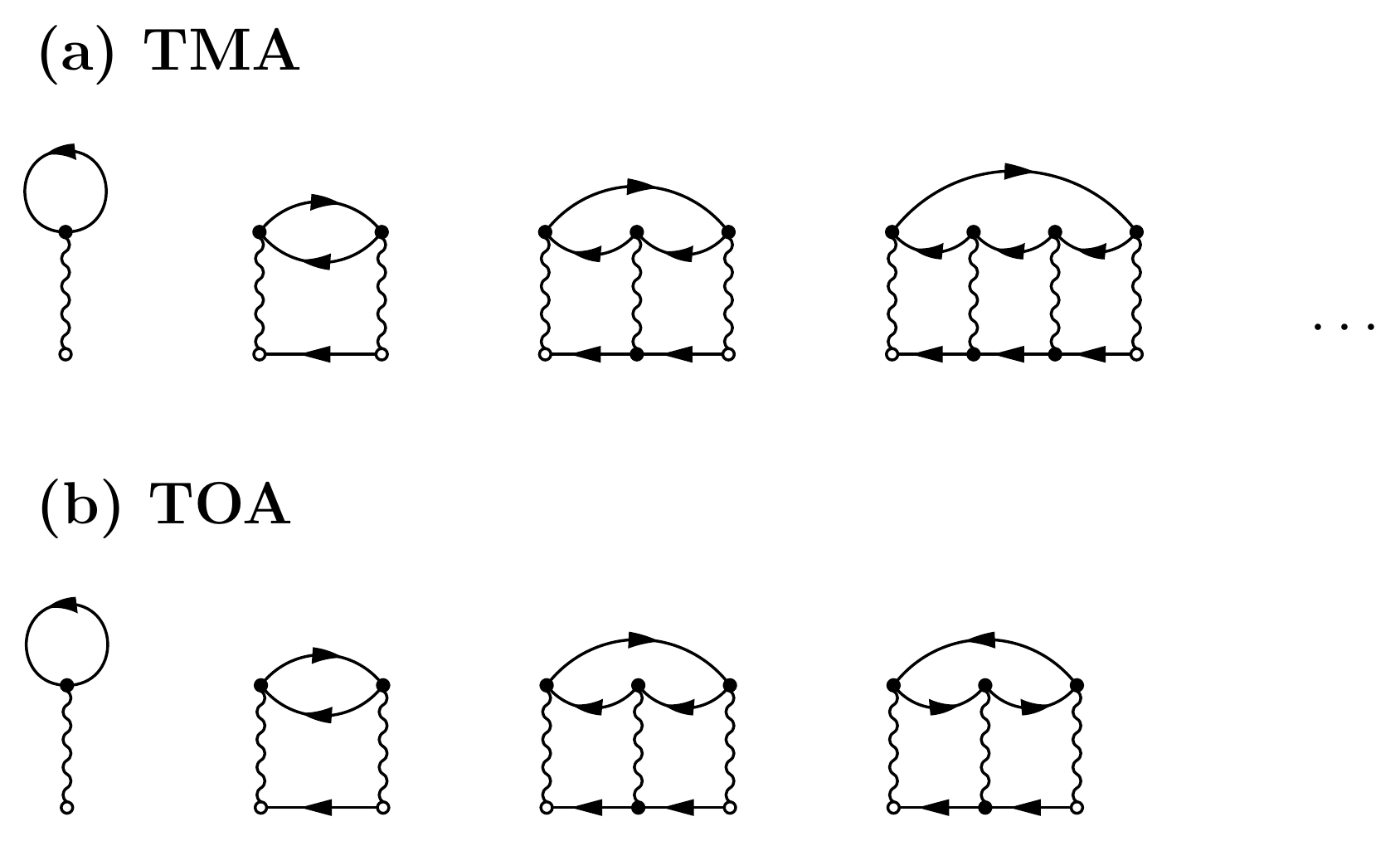}
\caption{
Feynman diagrams of the considered selfenergy approximations: (a) Diagram series of the particle-particle $T$-matrix approximation (TMA). (b) Diagrams contained in the third-order approximation (TOA), see text.
}\label{fig:diagrams}
\end{figure}

In the following we list the selfenergies that are used in the present paper. 
The contribution of the first order in the interaction is given by the Hartree--Fock (mean field) selfenergy,
\begin{align}
  \S{\sxbm}{\spbm}{\tn{HF},\uparrow(\downarrow)}(z,z') = U \dc (z-z') \kronecker{\sxbm}{\spbm} \n{\sxbm}{\downarrow(\uparrow)}(z)\, ,
\end{align}
which is contained in each of the approximations used below.
Many-body approximations that go beyond the mean field level (that are of higher than first order in $U$) contain, in addition, a correlation selfenergy, i.e.,
 ${\S{\sxbm}{\spbm}{\Sx} =: \S{\sxbm}{\spbm}{\tn{HF},\Sx} + \S{\sxbm}{\spbm}{\tn{cor},\Sx}}$. 
 %where in $\S{\sxbm}{\spbm}{\tn{cor},\Sx}$ higher order correlations can be included.
The first correlation correction is of second order and works well for weak coupling, i.e. $U \lesssim J$, for a discussion see Ref.~\cite{hermanns_prb_14}. Here we want to go beyond the weak coupling regime. Therefore, we focus on two higher order many-body approximations. The first is the  $T$-matrix approximation (TMA) in the particle-particle channel and yields a selfenergy $\S{\sxbm}{\spbm}{\tn{cor},\Sx}$ which accounts for scattering processes up to infinite order (see Ref.~\onlinecite{schluenzen_cpp_16} for a detailed discussion). This is realized by the selfconsistent, recursive structure of the $T$-matrix which can be understood as an effective interaction that obeys its own equation of motion (the Lippmann-Schwinger equation), Eq.~(\ref{eq:lsg}),
\begin{align}
 \S{\sxbm}{\spbm}{\tn{TMA},\uparrow(\downarrow)}(z,z') &= \i\hbar\, \T{\sxbm}{\spbm}(z,z')\,\G{\spbm}{\sxbm}{\downarrow(\uparrow)}(z',z)\,, 
\label{eq:sigma_t}\\
 \T{\sxbm}{\spbm}(z,z') &= -\i\hbar\, U^2\, \G{\sxbm}{\spbm}{\uparrow}(z,z')\, \G{\sxbm}{\spbm}{\downarrow}(z,z') \label{eq:lsg} \\
 &+\i\hbar\, U \intc{\cbar{z}} \G{\sxbm}{\sbbm}{\uparrow}(z,\cbar{z})\, \G{\sxbm}{\sbbm}{\downarrow}(z,\cbar{z})\T{\sbbm}{\spbm}(\cbar{z},z')\,. \nonumber
\end{align}
The corresponding Feynman diagrams are shown in \fref{fig:diagrams}(a). The TMA is known to perform best in the limit of small (or large) density~\cite{schluenzen_cpp_16, friesen09, hermanns_phd}, i.e., when the interaction in the system is dominated by electron-electron or hole-hole scattering events. Around  half-filling, however, electron-hole scattering gains in importance which is not captured by the particle-particle TMA. Therefore, we introduce, in addition, the third-order approximation (TOA) which contains all selfenergy contributions up to $\mathcal{O}\left( U^3 \right)$. In this approximation the correlation selfenergy, $\S{\sxbm}{\spbm}{\tn{cor},\Sx}$, attains the following form~\cite{hermanns_phd},
\begin{align}
  \S{\sxbm}{\spbm}{\tn{TOA},\uparrow(\downarrow)}(z,z') = -&\left(\i\hbar\right)^2\, U^2\\ & \nonumber \G{\sxbm}{\spbm}{\uparrow}(z,z')\, \G{\sxbm}{\spbm}{\downarrow}(z,z')\,\G{\spbm}{\sxbm}{\downarrow(\uparrow)}(z',z) \\
 -&\left(\i\hbar\right)^3\, U^3 \intc{\cbar{z}} \G{\sxbm}{\sbbm}{\uparrow}(z,\cbar{z})\, \G{\sxbm}{\sbbm}{\downarrow}(z,\cbar{z}) \nonumber \\ & \G{\sbbm}{\spbm}{\uparrow}(\cbar{z},z')\, \G{\sbbm}{\spbm}{\downarrow}(\cbar{z},z')\,\G{\spbm}{\sxbm}{\downarrow(\uparrow)}(z',z) \nonumber \\
 -&\left(\i\hbar\right)^3\, U^3 \intc{\cbar{z}} \G{\sxbm}{\sbbm}{\uparrow(\downarrow)}(z,\cbar{z})\, \G{\sbbm}{\sxbm}{\downarrow(\uparrow)}(\cbar{z},z) \nonumber \\ & \G{\sbbm}{\spbm}{\uparrow(\downarrow)}(\cbar{z},z')\, \G{\spbm}{\sbbm}{\downarrow(\uparrow)}(z',\cbar{z})\,\G{\sxbm}{\spbm}{\downarrow(\uparrow)}(z,z') \nonumber \, .
\end{align}
The corresponding diagrams are shown in \fref{fig:diagrams}(b). In the TOA, particle-particle and electron-hole scattering processes are considered on equal footing, yet only to third order inclusively. Both the TMA and TOA approach have been found to perform well for weak to moderate coupling strengths as long as the respective density conditions are fulfilled~\cite{schluenzen_cpp_16,friesen09,hermanns_phd,schluenzen16,hermanns_prb_14,bonitz_cpp_15,friesen09,friesen10,balzer_prb_16}.

Finally, we introduce the generalized Kadanoff-Baym ansatz (GKBA) which is an approximation that reduces the complexity of the time structure of NEGF theory by separating the time-diagonal Green functions from the off-diagonal ones. The full KBE of \eref{eq:kbe} is solved only for ${z=z'}$, while, for ${z\neq z'}$, the Green function is reconstructed from its time-diagonal values, i.e., from the single-particle density matrix. For the latter, the less and greater component of $G$ which originate from the mapping of the time contour onto the real time axis~\cite{balzer13} are reconstructed according to~\cite{lipavski},
\begin{align}
   \G{\sxbm}{\spbm}{\gtrless,\Sx}(t,t') \approx - & \left[ \G{\sxbm}{\sbbm}{\tn{R},\Sx}(t,t') \mathfrak{n}^{\gtrless,\Sx}_{\sbbm \spbm}(t') \right. \label{eq:GKBA} \\
   - & \left. \mathfrak{n}^{\gtrless,\Sx}_{\sxbm \sbbm}(t) \G{\sbbm}{\spbm}{\tn{A},\Sx}(t,t') \right]\, , \nonumber
  \end{align}
where ${\mathfrak{n}^{<,\Sx}_{\sxbm \spbm}(t) = \spdm{\sxbm}{\spbm}{\Sx}(t)}$ and ${\mathfrak{n}^{>,\Sx}_{\sxbm \spbm}(t) = \spdm{\sxbm}{\spbm}{\Sx}(t) - \kronecker{\sxbm}{\spbm}}$. The GKBA does not violate the attractive properties of the NEGF method, as it retains density and energy conservation, as well as time reversibility \cite{scharnke_17}. 
When using the GKBA, still the question remains how the  retarded and advanced propagators $G^{\tn{R}/\tn{A}}$~\cite{latini} are approximated. Here, we concentrate on Hartree--Fock propagators---the resulting approximation will be called HF-GKBA~\cite{hermanns_prb_14}. This approximation has been shown to eliminate (or drastically reduce) the artificial damping properties of two-time simulations for strongly excited systems and, at the same time substantially improving the computational performance.  
\subsection{Time-dependent density matrix renormalization group method (DMRG)}
\label{sec:tdmrg}
The density matrix renormalization group method \cite{white92,schollwoeck05,schollwoeck11} relies on approximating many-body wave-functions $|\psi\rangle $ via matrix-product states
of a finite bond dimension $m$.
A matrix-product state can be written as
\begin{equation}\label{eq:mps}
|\psi \rangle = \sum_{\sigma_1 \dots \sigma_L} A^{\sigma_1} A^{\sigma_2} \dots  A^{\sigma_L} |\sigma_1 \dots \sigma_L\rangle\,,
\end{equation}
where $\sigma_\ell$ are the local degrees of freedom at site $\ell$ and $A^{\sigma_\ell} $ are matrices of dimensions $m \times m$ (for details and the
role of boundary conditions, see Ref.~\onlinecite{schollwoeck11}).
Any wave-function $|\psi \rangle$ can be brought into the form Eq.~\eqref{eq:mps} by a sequence of singular value decompositions where in general, the bond dimension
of the matrices will scale exponentially in system size.
To illustrate this procedure, consider a one-dimensional system that is cut into two parts $A$ and $B$. By denoting complete basis sets in the parts $A$ and $B$ 
by ${|a\rangle_A}$ and ${|b\rangle_B}$, we can express a many-body wave-function as 
\begin{equation}
|\psi\rangle = \sum_{a,b} \psi_{a,b} |a\rangle_A  |b\rangle_B\,.
\end{equation}
By means of a singular value decomposition
of the rectangular matrix $\psi_{a,b}$, this can be reexpressed in terms of new basis sets in $A$ and $B$  
with a single index $\alpha$
\begin{equation}
|\psi\rangle = \sum_{\alpha=1}^s s_{\alpha} |\alpha\rangle_A  |\alpha\rangle_B
\end{equation}
where the $s_{\alpha}$ are the singular values and $s$ is the Schmidt number, which in general scales  exponentially with system size.
At the heart of the approximation used in DMRG and matrix-product states methods in general is a truncation in the number of states 
used to represent $|\psi \rangle$
by keeping only those $m$ states $ |\alpha\rangle_A  $ with the largest Schmidt coefficients $s_1^2  \geq s_2^2 \geq s_m^2 \geq \dots \geq s_s^2$, i.e.,
\begin{equation}
|\psi\rangle \approx  \sum_{\alpha=1}^m s_{\alpha} |\alpha\rangle_A  |\alpha\rangle_B\,.
\label{eq:trunc}
\end{equation}
This is equivalent to diagonalizing the reduced density matrix of part $A$
and truncating in its eigenspectrum, which was White's original formulation \cite{white92}
\begin{equation}
\rho_A = \mbox{tr}_B |\psi\rangle \langle \psi| = \sum_\alpha s_\alpha^2  |\alpha\rangle_A {}_A\langle \alpha |\,.
\end{equation}

While actual algorithms are described comprehensively in Ref.~\onlinecite{schollwoeck11}, we here want to explain 
for which many-body states Eq.~\eqref{eq:trunc} provides a useful approximation in the sense that
few states (order of $m\sim 100,1000$) suffice to obtain numerically accurate results for observables $\langle \hat O \rangle = \langle \psi | \hat O | \psi\rangle$.
This obviously depends on how quickly the eigenvalues $s_\alpha^2$ of the reduced density matrix decay.
A correct intuition can be gained from relating the decay of $s_\alpha^2$ to the entanglement entropy
\begin{equation}
S_{\rm vN} = - \mbox{tr}[ \rho_A \mbox{log} \rho_A] = - \sum_{\alpha} s_\alpha^2 \mbox{log}s_\alpha^2\,.
\end{equation}
A fast decay of $s_\alpha^2$ translates into a weakly entangled wave-function and vice versa. The crucial question
is the scaling of the entanglement entropy with the system size. For ground-states of gapped Hamiltonians with short-range
interactions, an area law holds \cite{eisert10}
\begin{equation}
S_{\rm vN} \propto L_A^{D-1}\,,
\end{equation}
where $L_A$ is the linear dimension of subsystem $A$
and $D$ is the dimension. Scaling in $L_A$ translates directly into scaling in $L$, i.e., the liner dimension of the full system.
In $D=1$, we obtain $S_{\rm vN} =\mbox{const}$ for $L \gg \xi$ where $\xi$ is the correlation length and this implies that the 
numerical effort (i.e., the number of states $m$ used to approximate $|\psi\rangle$) does not increase with system size 
since $m \lesssim \mbox{exp}(S_{\rm vN}(L))$ \cite{schollwoeck11}.
For critical systems in one dimension, the entanglement entropy acquires a logarithmic correction (see Refs.~\onlinecite{schollwoeck11,eisert10} and references therein).
This reasoning explains why matrix-product states based techniques work primarily for one-dimensional systems since in 2D, even if an area law holds, the
scaling is exponential in the width of the system \cite{Stoudenmire2012}.
The other important issue is whether an efficient algorithm can be formulated based on matrix-product states. It turns out that most matrix-product state methods including DMRG
scale as $m^3$ and linearly in $L$ \cite{schollwoeck11}.

For real-time evolutions $|\psi(t) \rangle = \mbox{exp}(- \i H t) |\psi(t=0) \rangle$, the application of the time-evolution operator
can be efficiently implemented via a Trotter-Suzuki decomposition (using $H=\sum_\ell h_{\ell,\ell+1}$) into operators $\mbox{exp}(-\i h_{\ell,\ell+1}\delta t)$
which is just a local two-site gate affecting two $A$-matrices in Eq.~\eqref{eq:mps} ($\delta t$ is the time step) \cite{Vidal04,daley04,white04}.
In general, a time-propagated many-body state $|\psi(t)\rangle$ will develop volume-law like entanglement even if the initial state was a product state \cite{schollwoeck11}.
For global quantum quenches (such as the time-evolution from a product state such as our case (iv), see Fig.~\ref{fig:initial_states}(d)), the entanglement 
grows linearly in time $S_{\rm vN} \propto t$, implying that the number of states $m$ needed to maintain the same quality of approximation to the true 
$|\psi(t)\rangle$ will increase exponentially. This limits the accessible times in global quenches to about $t\lesssim \mathcal{O}(10/J)$, while in local quenches, geometric quenches such as the sudden expansion considered in examples (i)-(iii)  or for slow parameter changes, a milder entanglement increase occurs.

The two main parameters that  control the accuracy of time-dependent DMRG simulations are the time step $\delta t$ and the discarded weight $\delta \rho$ \cite{gobert05,schollwoeck11}.
The latter is defined as 
\begin{equation}
\delta \rho = \sum_{\alpha=m+1}^s s_{\alpha}^2\,,
\end{equation}
 which is a measure for the error made per truncation. The quality of DMRG data has to be analyzed as a function of both $\delta t$ and $\delta \rho$, 
with the latter the dominant parameter since the dependence of the error on $\delta t$ can be reduced by using higher-order Trotter-Suzuki decompositions \cite{schollwoeck11}. In this work, we use a time-dependent DMRG implementation as introduced in Refs.~\onlinecite{white04,daley04} and we varied the time step between
$0.02/J \leq \delta t  \leq 0.1/J$ and the discarded weight $ 10^{-7} \leq \delta \rho \leq 10^{-4}$ with a maximum number of $2000$ states.

\section{Numerical Results}\label{s:results}
In our simulations we consider four different nonequilibrium setups. The corresponding initial states are depicted in \fref{fig:initial_states}. In all cases, the incipient configuration consists of both, doubly occupied and empty Hubbard sites. In the first setup, the occupied sites are arranged on a straight line to form a one-dimensional band insulator (BI). During time propagation, and in the absence of any further potential, the density  starts to expand symmetrically towards the left and right edges of the Hubbard chain, cf.~\fref{fig:initial_states}(a). Next, in order to also investigate an asymmetric expansion dynamics the initial BI is placed onto the leftmost sites of the chain allowing the density to escape  only to the right, see \fref{fig:initial_states}(b). Further, to analyze the effect of the dimensionality of the system we extend the asymmetric setup to a  two-leg Hubbard ladder the leftmost rungs of which are initially doubly occupied, \fref{fig:initial_states}(c). The dynamics on such ladders is often used to investigate the 1D-to-2D crossover. Finally, we consider a setup that generates a final state at a constant and large density  where correlation effects are expected to manifest themselves even stronger, cf.~\fref{fig:initial_states}(d). Here, the initial state consists of a one-dimensional Hubbard chain with alternating occupation $n_i =0,2$. During the evolution from this charge-density wave (CDW),  the particles quickly form an entangled many-body state in which correlations play a crucial role. These four setups will be analyzed in detail in Secs.~\ref{sec:exp_1D}--\ref{sec:cdw}.
%
%\subsection{Sudden expansion}
%\label{sec:sudden}
%
\subsection{Sudden expansion in 1D: symmetric case}
\label{sec:exp_1D}
We start the numerical analysis by considering a confinement quench giving rise to a sudden symmetric expansion of an ensemble of fermions into an empty lattice (see \fref{fig:initial_states}(a)). This setup has been studied in many papers, including experimental studies \cite{schneider12,ronzheimer13,xia14,vidmar15}, and theoretically using 
DMRG methods \cite{hm08,hm09,kajala11,langer12,bolech12,vidmar13,Mei2016,Hauschild2015} and 
NEGF \cite{schluenzen16}. 
We exclusively study an initial density of $n=2$  on the sites that are occupied at $t=0$, which was previously considered in Refs.~\onlinecite{kajala11,kessler13,schluenzen16}
such that the general properties are well understood.

\begin{figure}[tb]
\includegraphics[width=\columnwidth]{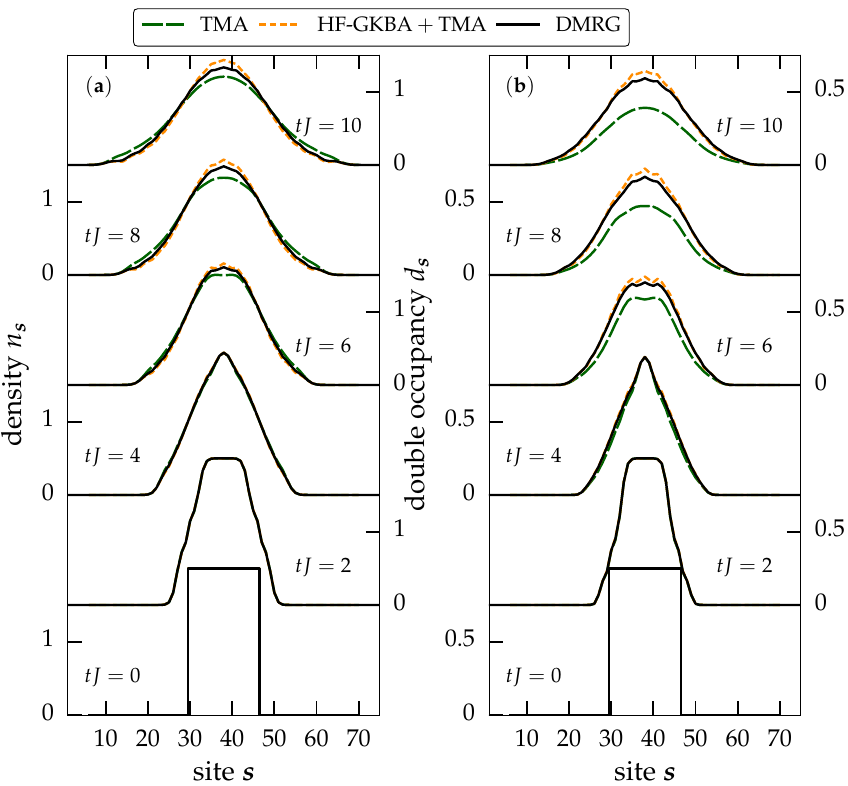}
\caption{(Color online)
Symmetric 1D sudden expansion of a Hubbard chain of $N=34$ fermions at $U=J$. Time evolution of (a) density $n_{\bm{s}}$ and (b) double occupancy  $ d_{\bm{s}}$ for 6 times (from bottom to top): $tJ=0,2,4,6,8,10$. Solid lines: DMRG, dashes: NEGF (two-time $T$-matrix), Dotted lines: $T$-matrix with HF-GKBA.
\label{fig:ni_d}
}
\end{figure}

%In the first setup, the time propagation starts with a finite ensemble of particles initially confined at the center of a one-dimensional Hubbard chain, cf.~\fref{fig:initial_states}(a). 
%This kind of setup has already been investigated separately using the NEGF method\cite{schluenzen16} as well as the DMRG\cite{kajala11} and the general properties are well understood. 
Here, we focus on the quantitative details of the time dynamics and compare our NEGF results to DMRG. We consider a chain of length $L=75$ with $N=34$ particles for $U=J$. The evolution of the respective density profiles is shown in \fref{fig:ni_d}(a) for six consecutive time steps $tJ = 0,2,4,6,8,10$. 
The solid, black lines correspond to the DMRG results and the dashed green lines belong to the NEGF calculations using the $T$-matrix approximation (TMA) while the orange lines are obtained by additionally invoking the HF-GKBA, cf. \sref{sec:negf}. As expected, the general trend of the density is to propagate outwards resulting in a bell-shaped profile which can be seen from all considered descriptions. For times exceeding $5 J^{-1}$, the site occupations start to deviate slightly in the three simulations. In the full two-time NEGF calculation the fermion expansion is slightly faster than in the DMRG, while in the HF-GKBA simulation the particles stay closer to the center and are in very good agreement with the DMRG. 

A quantity more sensitive to correlations is the double occupancy, 
\begin{eqnarray}
 \doc{\sxbm}{} := \n{\sxbm}{\uparrow\downarrow} = \avg{\cop{\sxbm}{\uparrow}\aop{\sxbm}{\uparrow}\cop{\sxbm}{\downarrow}\aop{\sxbm}{\downarrow}}\, ,
\end{eqnarray}
the dynamics of which are displayed in \fref{fig:ni_d}(b). It is evident that it follows the trend of the density by which it is dominated. Again, in the full two-time NEGF calculation $\doc{\sxbm}{}$ expands faster than in the DMRG result, where the deviations are larger than for the density. In contrast, the HF-GKBA is again very close to the latter.

\begin{figure}[tb]
\includegraphics[width=\columnwidth]{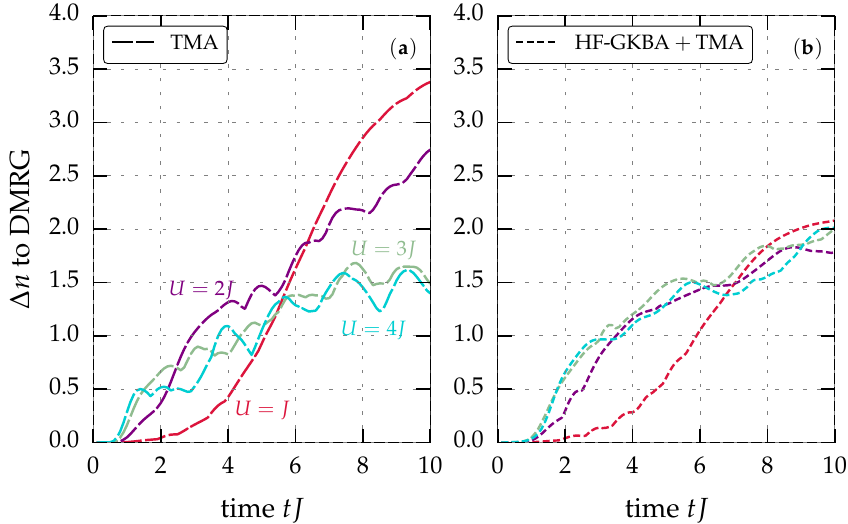}
\caption{(Color online)
Total density deviation, \eref{eq:dn}, between (a) DMRG and NEGF and (b) DMRG and HF-GKBA, for the symmetric 1D sudden expansion. Simulations as in \fref{fig:ni_d} with $N=34$ particles, but for four different values of $U$ indicated in the figure.
%\fabian{Perhaps use symbols?}
}\label{fig:dev_time}
\end{figure}

To better quantify the discrepancies between DMRG and the two NEGF approaches we introduce the total density deviation between the two methods in the following way,
\begin{eqnarray}
 \Delta n(t) := \sum_{\sxbm} \left| \n{\sxbm}{\tn{I}}(t) - \n{\sxbm}{\tn{II}}(t) \right|\, , \label{eq:dn}
\end{eqnarray}
where $\tn{I}$ and $\tn{II}$ denote the respective method. This quantity allows to analyze the time dependent difference of the density profiles. It should be noted that the quantitative value of $\Delta n$ has no direct interpretation. Instead, by dividing by the total number of Hubbard sites $\nsx$, one gets the average deviation per site. Adopting DMRG as the reference method, we investigate the dependence of the deviation on the interaction strength $U$ and time by calculating the total deviations for the two-time TMA simulation and the HF-GKBA results which are displayed in \fref{fig:dev_time}(a) and (b), respectively.
As expected, the total deviation grows in time for all cases. Interestingly, however, the deviations saturate around $t = 10 J^{-1}$. A closer look reveals that, during the early propagation period, the growth appears to be superlinear, followed by a receding phase after which the growth becomes more fluctuating. The lengths of these time periods strongly depends on the interaction strength, as they become shorter for larger $U$. As a consequence, for times around $tJ=1$ the total deviation increases with the interaction strength while for times around $tJ=9$ it decreases with $U$. The overall trend is common between the full TMA results and the HF-GKBA simulations. The only noticeable difference is that $\Delta n$ remains a little smaller for larger times and small $U$ in the HF-GKBA calculations. 

\begin{figure}[tb]
\includegraphics[width=\columnwidth]{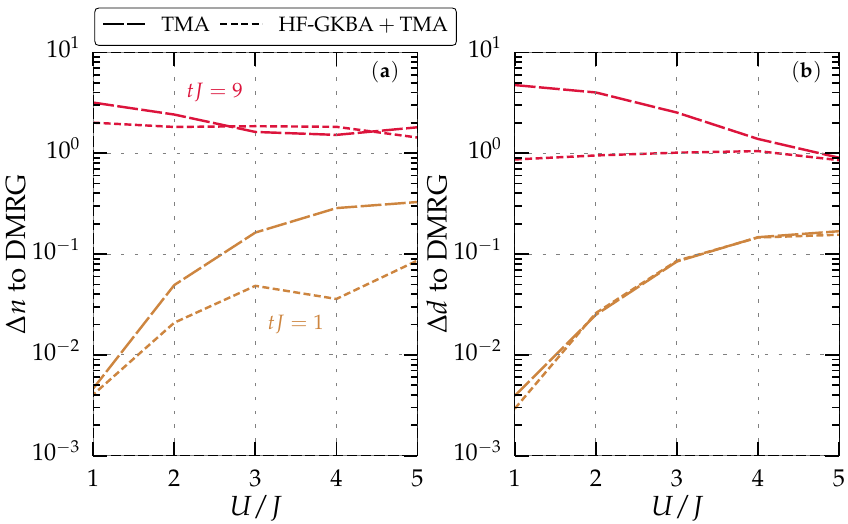}
\caption{(Color online)             
{\it Symmetric 1D sudden expansion.} Total deviation of (a) the density  and  (b) the double occupancy  between DMRG and NEGF (solid lines) and DMRG and HF-GKBA (dashed lines) at $tJ=1$ and $tJ=9$,
%versus $U/J$ 
for $N=34$. 
%(b) Total deviation in the double occupancy between DMRG and NEGF (solid lines) and DMRG and HF-GKBA (dashed lines) at $tJ=1$ and $9$
%versus $U/J$ ($N=34$, $L_0=17$). 
}\label{fig:dev_U}
\end{figure}

To better understand how the density deviations vary with $U$ and $t$, we replot these quantities in \fref{fig:dev_U}(a) for the two time points, $tJ=1$ and $tJ=9$. In addition, we compute the total deviation of the double occupancy which is defined, by analogy to \eref{eq:dn}, as
\begin{eqnarray}
\Delta d(t) := \sum_{\sxbm} \left| \doc{\sxbm}{(1)}(t) - \doc{\sxbm}{(2)}(t) \right|\, ,
\end{eqnarray}
which is shown in \fref{fig:dev_U}(b).  
While for $tJ=1$ all results confirm the trend that the deviations grow with increasing $U$, for $tJ=9$ the dependence is more irregular. In the latter case, the deviations between HF-GKBA and DMRG, in particular, are nearly independent of the interaction strength. The two-time NEGF results for the double occupancy, however, show large deviations for small $U$. The decrease of $\Delta n$ and $\Delta d$ with increasing $U$ for later times can be understood from the direct dynamics of the density profiles. Since  for large $U$ the particles predominantly remain in the center of the system the growth of the deviations is limited due to the absence of moving particles. 

From the presented results, it turns out that the DMRG result is typically enclosed between the HF-GKBA and the two-time NEGF result in $T$-matrix approximation.
At the same time the HF-GKBA data are slightly closer to the DMRG results. 

\begin{figure}[tb]
\includegraphics[width=\columnwidth]{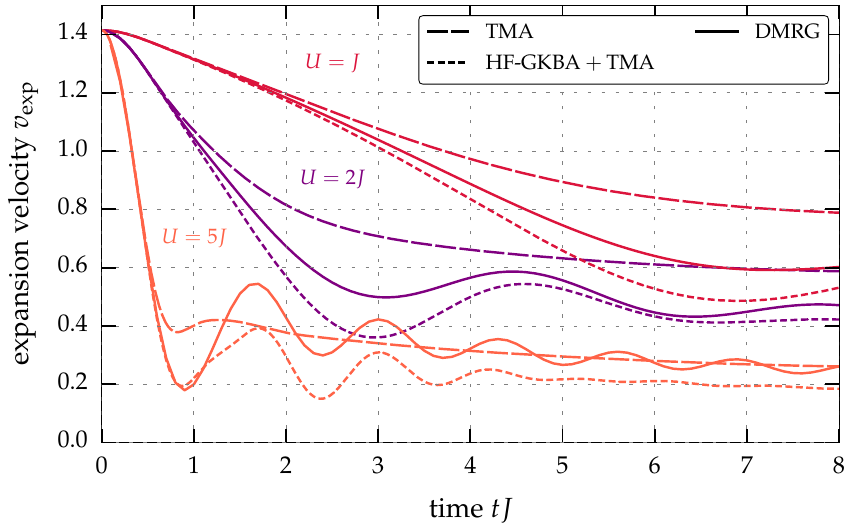}
\caption{(Color online)
Symmetric 1D sudden expansion.
Evolution of the expansion velocity versus time for $U/J=1,2,5$ for three simulation methods.
}\label{fig:vexp}
\end{figure}

To further analyze the expansion behavior following a 1D sudden confinement switch, it is instructive to analyze 
%a quantity that is little more involved in the dynamical properties of the evolution. For this purpose one can introduce 
the expansion velocity of the fermion cloud which is defined according to~\cite{schluenzen16}
\begin{align}\label{eq:vexp_and_d}
 v_{\tn{exp}}(t)&= \frac{\d}{\d t} D(t), \, \tn{with} \quad D(t) = \sqrt{R^2(t)-R^2(0)} \, , \\ \nonumber
  R^2(t) &= \frac{1}{N} \sum_{\sxbm} \n{\sxbm}{} (t)\, \lVert\sxbm-\sxbm_0\rVert^2 \,, 
\quad \sxbm_0 = \frac{1}{N} \sum_{\sxbm} \n{\sxbm}{} (0)\,\sxbm \,.
\end{align}
This quantity measures the temporal growth of the particle cloud which has a mean square radius $R(t)$ from which the initial size is subtracted. This quantity was analyzed in detail for 1D, 2D and 3D systems and a broad range of system size $N$ in Ref.~\onlinecite{schluenzen16}.
Here we focus on the time evolution of $v_\tn{exp}$ for 1D systems and compare again DMRG, two-time NEGF simulations and HF-GKBA. The results are shown in \fref{fig:vexp} for $U/J=1,2,5$. 
As one can see, for all cases $v_\tn{exp}$ starts from the same value $v_\tn{exp}(0)=v_\tn{max} = \sqrt{2D}J = \sqrt{2}J$ which is the largest 
 expansion velocity 
in an empty lattice (cf.~Refs.~\onlinecite{schneider12,langer12,schluenzen16}). 
A noninteracting gas expands with a a constant $v_\tn{exp}(0)=v_\tn{max} = \sqrt{2D}J$ \cite{schneider12}.
For $U>0$,  $v_\tn{exp}$ decreases from its initial value until it slowly converges to an approximately  constant value once $U/J$ becomes comparable to the bandwidth.
 
This behavior is explained by the large effective mass of doublons in the limit $U\gtrsim 4J$, where perturbation theory results in an effective hopping matrix elements
$J_{\rm doublon} \propto J^2/U$ for $U\gg J$ (see, e.g., Ref.~\onlinecite{winkler06}). As a consequence, the doublons become inert on the accessible time scales 
and the system remains largely in a weakly-correlated state that is essentially a product state in the core region \cite{hm09,kajala11,bolech12} (see also Refs.~\onlinecite{Hofmann2012,muth12,Rausch2016}).
Therefore, $v_\tn{exp}$ is dominated by the few atoms that expand after some doublons have dissolved into single particles \cite{kajala11} and at long times, the
expansion velocity is dominated by these fast atoms while the slow doublons do not contribute \cite{vidmar13}.

For the applicability of the NEGF methods, this dynamical freezing of a build-up of correlations as $U/J \gtrsim 4$ implies that the methods become
more accurate again, since the wave-functions acquire a simpler structure than at weak $U/J$. This explains the a priori counterintuitive observation that
the numerical deviations of the NEGF methods compared to DMRG (see the discussion of Fig.~\ref{fig:dev_time} and Fig.~\ref{fig:dev_U}) become smaller as $U/J$ increases, even though the
NEGF techniques are by construction weak-coupling methods.  
The regime of $U/J\gg 4$ can much easier be accessed by DMRG with longer times becoming accessible \cite{hm09}, demonstrating the 
usefulness of NEGF and DMRG as complementary approaches for weak and strong coupling, respectively.

Interestingly, for large interaction strength both DMRG and the HF-GKBA propagation show oscillations in the expansion velocity with similar frequency. In contrast, for the two-time TMA calculations an onset of oscillations is seen only for $U/J=5$, in all other cases the expansion velocity quickly damps monotonically approaching an asymptotic value. 
This is consistent with earlier observations that two-time propagations of the KBEs for strongly excited small Hubbard systems can be accompanied by
 unphysical damping in the density evolution \cite{friesen09,friesen10,hermanns_prb_14}, as was noted in the introduction Sec.~\ref{s:intro}. Since the initial confinement quench in our simulations constitutes such a strong excitation it is very likely that the missing of the oscillations of the expansion velocity in the two-time simulations are associated with this artificial damping. 

On the other hand, the HF-GKBA is known to remove the artificial damping in strongly excited small systems \cite{hermanns_prb_14}. Therefore, it is not surprising, that in the present setup, the HF-GKBA simulations exhibit better agreement with the DMRG for intermediate times, including the reproduction of the oscillations of the expansion velocity. This is particularly the case for small and moderate couplings, $U \lesssim 3J$. For larger couplings, the long-time asymptotics of the expansion velocity of the two-time simulations is closer to the DMRG than the HF-GKBA result.
% 
%From this, it becomes apparent that the HF-GKBA is more suited to describe the dynamics of the system, including also the fluctuations on a shorter time scale which also explains the better performance in 
This behavior is also consistent with the earlier observations for the evolution of the density profile and the double occupancy. 
This complementarity of the performance of the two-time and the HF-GKBA NEGF simulations are a particularly attractive feature. 

Therefore, having 
both NEGF results at hand, allows one to  estimate, e.g., the value of the asymptotic expansion velocity. For all $U$, the DMRG solution of this asymptotic value lies within the NEGF results. Utilizing this observation, one can extract the exact value of $\lim_{t \to \infty}v_\tn{exp}(t)$ with a relative error of $\lesssim 30 \%$, for all $U$. 

The experiment \cite{schneider12} used a different measure for the expansion velocity derived from the time evolution
of the half-width-at-half-maximum, called core expansion velocity.
In Ref.~\onlinecite{schluenzen16}, a direct comparison of numerical results for this core expansion velocity to experimental data of Ref.~\onlinecite{schneider12}
was presented, with a very good agreement. 
Our analysis of the errors of densities as a function of $U/J$ and time in the different NEGF schemes  further corroborates the
validity of the NEGF data used in that comparison.

We close this discussion by noting that in principle, it should be possible to compute the asymptotic expansion velocities from the Bethe ansatz, along the 
lines of Refs.~\onlinecite{bolech12,Mei2016}. For instance, $\lim_{t \to \infty}v_\tn{exp}(t)$ was computed for $n\leq 1$ with an excellent agreement with DMRG results \cite{Mei2016}.  The extension of Ref.~\onlinecite{Mei2016} to initial densities $n \gtrsim 1$ is left for future research. 

\subsection{Sudden expansion in 1D: asymmetric case}\label{sec:asym}
\begin{figure}[tb]
\includegraphics[width=\columnwidth]{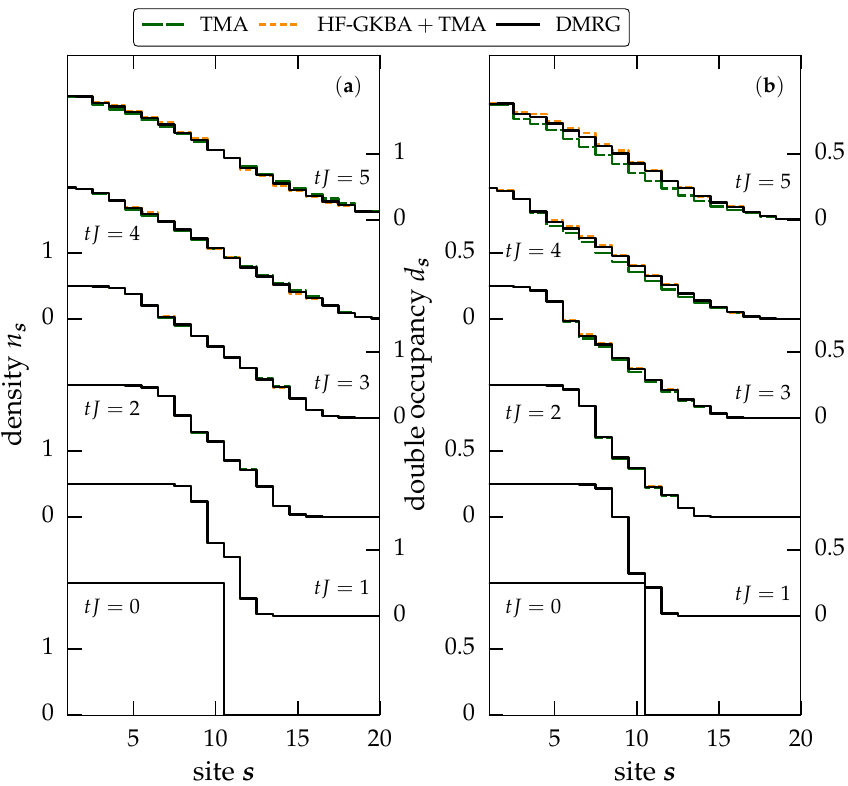}
\caption{(Color online)
{\it Asymmetric 1D sudden expansion.} Time evolution of (a) density $ n_s $ and (b) double occupancy  $ d_s$
for $U=J$, $N=20$ for $tJ=0,1,2,3,4,5$. Solid lines: DMRG, dashes: NEGF (two-time $T$-matrix), Dotted lines: $T$-matrix with HF-GKBA.
}\label{fig:asym_ni_d}
\end{figure}

\begin{figure}[tb]
\includegraphics[width=\columnwidth]{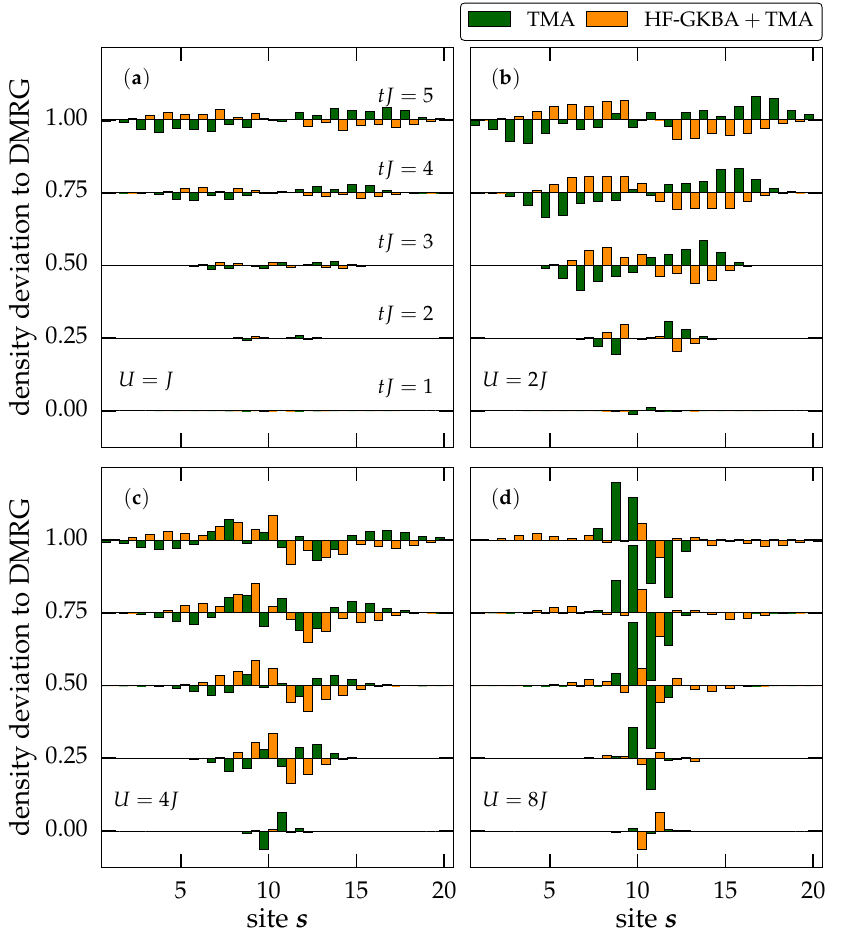}
\caption{(Color online)
Asymmetric 1D sudden expansion for $N=20$ fermions, initially doubly occupying the $L_0=10$ leftmost sites labeled $0, \dots, 9$. Total system length is 30 sites, 
cf. \fref{fig:initial_states}(b).
Deviation of the density profiles between DMRG and NEGF and DMRG and HF-GKBA for different time steps and (a) $U=J$, (b) $U=2J$, (c) $U=4J$, (d) $U=8J$.
Successive times are (from bottom to top) $tJ=1, 2, 3, 4, 5$ (each curve is vertically shifted by $0.2$). For better visibility, the HF-GKBA curves are horizontally displaced by one site.
The data for $tJ= 2, 3, 4, 5$ are vertically shifted by increments of $0.2$ for better visibility.
}\label{fig:profiles1}
\end{figure}

It is now interesting to further investigate whether the observed accuracy and the complementary behavior of single-time and two-time NEGF simulations is just a special case of the symmetric expansion. To this end we now consider a modified setup [cf. \fref{fig:initial_states}(b)] where the confinement quench gives rise to a density expansion in only one direction. The results are presented in \fref{fig:asym_ni_d}(a) for $N=20$ fermions and $U=J$. The respective evolution of the double occupancy is shown in \fref{fig:asym_ni_d}(b). As one can see, the results obtained by all considered methods lie very close to each other. To better distinguish between the particular profiles, we show the deviations to the DMRG results in \fref{fig:profiles1}. The subfigures \fref{fig:profiles1}(a)-(d) correspond to different interaction strengths $U/J=1,2,4,8$. 
%For $N=16$ fermions, the respective results are depicted in \fref{fig:profiles2}.
%
%
% \begin{figure}[tb]
% \includegraphics[width=\columnwidth]{FHM_density_profile_asymm_8.pdf}
% \caption{(Color online)
% Asymmetric 1D sudden expansion.
% Deviation of the density profiles between DMRG and NEGF and DMRG and HF-GKBA versus time for (a) $U=J$, (b) $U=2J$, (c) $U=4J$, (d) $U=8J$.
% Times $tJ=1,2,3,4,5$ for $L=20$, $N=16$, $L_0=8$.
% Same as \fref{fig:profiles1}, but for $N=16$ fermions, initially doubly occupying the $L_0=8$ leftmost sites. \fabian{How much do we really learn from showing 
% both figures 8 \& 9?}
% }\label{fig:profiles2}
% \end{figure}
As in the symmetric 1D setup before we observe a complementary behavior of the two-time result and the HF-GKBA. First, it is striking that again both approximations exhibit opposite deviations from the DMRG: while the two-time results show a slightly too fast expansion, the HF-GKBA results are retarded. Correspondingly, the deviations of the local densities from the DMRG results have  opposite signs: the HF-GKBA (two-time TMA) densities are above (below) the DMRG result, on the originally doubly occupied sites and, vice versa, for the unoccupied sites. Also, the deviations have a similar dependence on the coupling strength as in the symmetric case, \sref{sec:exp_1D}: For the HF-GKBA that exhibits smaller density deviations than the two-time result for all considered $U$, the maximum deviation is found for intermediate coupling strengths ($2J \le U \le 4J$). In contrast, for $U = 8J$, the two-time result shows large deviations at the edge of the occupied region. 
%These observations are confirmed in the simulations of a slightly smaller system shown in \fref{fig:profiles2}.

\begin{figure}[tb]
\includegraphics[width=\columnwidth]{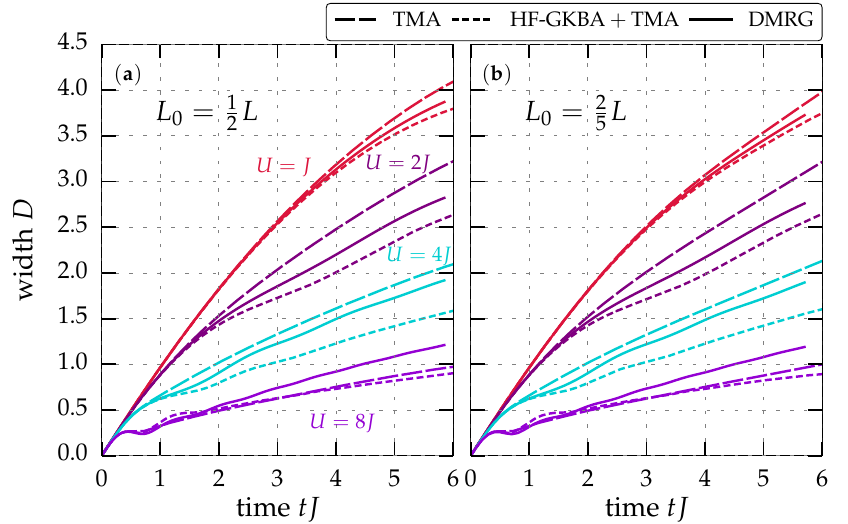}
\caption{(Color online)
Asymmetric 1D sudden expansion.
Reduced radius $D(t)$ defined by \eref{eq:vexp_and_d} as a function of time for four values of the interaction strength. (a): $L_0=L/2$, cf. \fref{fig:profiles1} and
(b): $L_0=2/5L$.
}\label{fig:width}
\end{figure}

After considering the densities we again compute the width of the expanding particle cloud, using \eref{eq:vexp_and_d} (see \fref{fig:width}(a)). This quantity confirms the observations made before for the density: compared to the DMRG, the expansion of the particle cloud is slightly accelerated (decelerated) for the two-time (GKBA) simulations. With increasing $U$ the two-time result becomes more accurate than the HF-GKBA. For large couplings, $U \gtrsim 6 J$, the deviations between two-time and single-time approximations vanish. In this limit, also the two-time result for the expansion is retarded, in comparison to the DMRG. 
This analysis indicates that for $U \lesssim 6J $ a combination of two-time and single-time simulations is able to reproduce the cloud size with a relative error not exceeding $20 \%$. However, for large couplings, the inaccuracies grow and appear to arise from the inadequacy of the underlying $T$-matrix approximation in the particle-particle channel. This makes it necessary to study additional many-body approximations. 

The considered system of $L=20$ sites and ${L_0=L/2=10}$ initially doubly occupied sites obeys a high symmetry between the electron- (hole-) density on site $\sxbm$ and the hole- (electron-) density on site $L-\sxbm$. To generalize our findings, we also present results for a system of ${L_0=2/5L=8}$ initially occupied sites, for which this symmetry is broken. The corresponding widths of the particle cloud are shown in \fref{fig:width}(b). As one can see, all trends agree with the previous results and especially the enclosing behavior of the NEGF methods seems not to depend on the symmetry of the system.

\subsection{Asymmetric sudden expansion on a two-leg ladder}\label{sec:ladder}
As mentioned before, the generalization to higher system dimensions constitutes a challenging problem to DMRG due to the additional degrees of freedom in the correlation growth. Therefore, only simple 2D toy models have been simulated so far with time-dependent DMRG, including the expansion of strongly interacting bosons 
on few-leg ladders and in small 2D clusters \cite{Hauschild2015}.
Since the NEGF method is not restricted with respect to dimensionality it is very interesting to compare the performance of both methods on a two-leg ladder to see if the good agreement of the previous 1D analysis can be confirmed for higher dimension.

\begin{figure}[tb]
\includegraphics[width=\columnwidth]{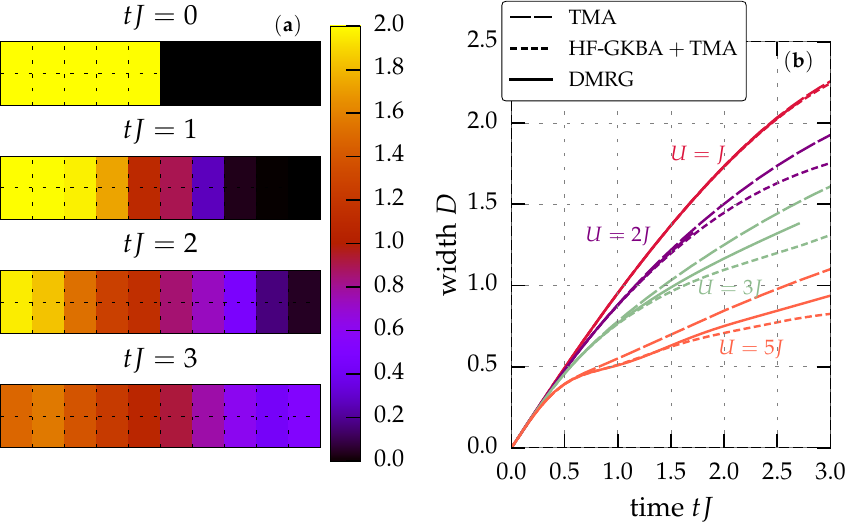}
\caption{(Color online)
Asymmetric sudden expansion on a two-leg ladder. (a) Density distribution for four timesteps $tJ=0,1,2,3$ of the HF-GKBA+TMA simulation for $U=J$. (b) Reduced radius $D(t)$ defined by \eref{eq:vexp_and_d} as a function of time for four values of the coupling parameter.
}\label{fig:ladder_width}
\end{figure}

As in \sref{sec:asym}, we consider an asymmetric expansion setup, now with a ladder of ten rungs, the leftmost five of which are initially doubly occupied (cf. \fref{fig:initial_states}(c)). The resulting density evolution is illustrated in \fref{fig:ladder_width}(a) where the density distributions for several time steps $tJ=0,1,2,3$ are shown in a simulation using the HF-GKBA+TMA approach with $U=J$. As in the 1D case, the particles tend to move to the right. To quantify the growth of the width of the particle cloud we again use the reduced radius $D(t)$ of \eref{eq:vexp_and_d} the time evolution of which is shown in \fref{fig:ladder_width}(b) for all considered methods and $U/J=1,2,3,5$. As one can see, the behavior is very similar to the 1D case (cf. \fref{fig:width}). The slowing down of the expansion for increasing interaction strength is well predicted by all considered methods for small interaction strengths, whereas for larger $U$, the DMRG curve lies between the NEGF results. 

It should be mentioned that the evolutions of the reduced radius for all $U$ share a common short-time phase (this is also present in 1D but becomes more apparent on the ladder), for which $D(t)$ behaves like the ideal system. This phase shortens with increasing interaction strength, which is due to the build-up of correlations. The behavior is similar for the symmetric expansion setup (cf. \sref{sec:exp_1D}) for which the dependence of the early expansion phases and the connection the onset of correlations are analyzed in detail in Refs.~\onlinecite{schluenzen_cpp_16,schluenzen16}.
\subsection{Relaxation of charge-density wave state of doublons}\label{sec:cdw}
We now turn to the fourth setup that is depicted in \fref{fig:initial_states}(d), an alternating sequence of doubly occupied and empty states corresponding to a charge-density wave. 
There have been a number of experiments starting from similar initial states with both fermions \cite{pertot14} or bosons \cite{trotzky12}, but mostly of the $|\psi_0 \rangle = |1,0,1,0, \dots \rangle $ type.
Theoretically, there is much interest in the decay of charge-density waves or initial states with perfect N{\'e}el order in the Fermi-Hubbard model, with previous
work on both its 1D version \cite{enss2012,Bauer2015} and for higher-dimensional systems (see, e.g., Ref.~\onlinecite{Balzer2015}). The decay from the bosonic version of 
our initial state $|\psi_0 \rangle = |2,0,2,0, \dots\rangle $ was studied in Ref.~\onlinecite{carleo12}. 

Here the dynamics is governed by a short-time process in which 
particles move into the empty sites, provided that $U <4J$, which is the band-width.
After that, a spreading and build-up of correlations sets in, for which the relevant velocity is typically
strongly dependent on $U/J$ \cite{cheneau12,enss2012,Bauer2015}.

\begin{figure}[tb]
\includegraphics[width=\columnwidth]{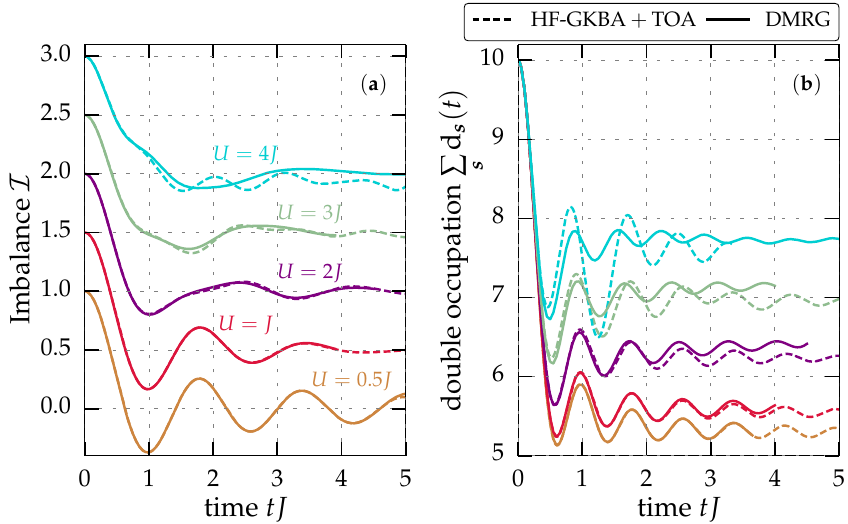}
\caption{(Color online)
{\it Relaxation of CDW state} of doublons for $U/J=0.5,1,2,3,4$ (each curve is vertically shifted by $0.5$) and $L=20$, $N=20$. 
(a) Doublon Imbalance, \eref{eq:imbalance}, 
(b) Total double occupancy $\sum_{\bm{s}} d_{\sxbm} (t)$.
}\label{fig:cdw}
\end{figure}

A useful quantity for the dynamics is the occupation imbalance which is defined as the difference of the densities on all even and all odd sites, 
\begin{align}
 {\cal I}(t) = \frac{N_\tn{even}(t)-N_\tn{odd}(t)}{L}\, ,
\label{eq:imbalance}
\end{align}
where $N_\tn{even}$ ($N_\tn{odd}$) sums up all densities of the even (odd) sites. The imbalance starts from $N/L$ and is then expected to decay. The results for $N=20$ fermions and five different couplings are shown in \fref{fig:cdw}(a). 
In the figure we compare DMRG results to NEGF simulations using a third-order approximation for the selfenergy. We show results for the single-time limit, i.e. after applying the HF-GKBA\footnote{For NEGF, we leave out the full two-time propagation results due to the artificial damping (cf. \fref{fig:vexp}) that completely cancels out all occurring oscillations in the CDW setup.}. The agreement is excellent for small and moderate couplings. Only once the interaction strength becomes as large as $U/J=3$,  small deviations are visible which grow for $U=4J$. This is not surprising because the third-order approximation does not capture higher order corrections. It contains, however, the third-order electron-hole diagram (cf. \fref{fig:diagrams}) which becomes essentially important at half filling. Therefore, TOA simulations are superior to the $T$-matrix calculations with respect to the description of CDW dynamics. Similar trends are seen in the total double occupancy which is displayed in \fref{fig:cdw}(b). 
The dependence of errors of the NEGF methods on $U/J$ (i.e., an increase as $U/J$ becomes order of the bandwidth)  in this example is the expected generic behavior since these methods
are by construction weak-coupling approaches. DMRG works particularly well for $U/J>4$ in such problems \cite{enss2012}, illustrating the 
complementary strength of NEGF versus DMRG time evolutions. 
% \begin{figure}[tb]
% %\includegraphics[width=\columnwidth]{PDF/fig_2D_N.pdf}
% \caption{(Color online)
% {\it Relaxation of CDW state of doublons.}
% Time evolution of quasi-MDF for $U=J$ and $U=4J$.
% $L=12$, $N=12$.
% \fabian{I think we will drop this, right?}
% }\label{fig:mdf}
% \end{figure}
% \begin{figure}[tb]
% \includegraphics[width=\columnwidth]{FHM_CDW_size_dualplot.pdf}
% \caption{(Color online)
% {\it Relaxation of CDW state of doublons.}
% System-size dependence for (a) $U=J$, (b) $U=4J$
% for $L=6,12,20,24,36$
% }\label{fig:Ldep}
% \end{figure}

\section{Summary and outlook}
\subsection{Summary of main results}

In this paper we thoroughly investigated the accuracy and applicability range of NEGF-based approaches in the description of the complex and correlated electron dynamics in strongly excited large Hubbard chains. The basis for this benchmark analysis were DMRG simulations performed for the same setups. Based on this analysis for the selected four setups we may conclude that NEGF simulations are reliable and accurate, thereby fully confirming earlier comparisons to exact diagonalization results for small clusters. Thus, NEGF simulations have predictive power, far beyond the present systems and situations. More precisely, our conclusions can be summarized as follows:
\begin{enumerate}
    \item The quality of the NEGF results crucially depends on the choice of the selfenergy, $\Sigma$, which is clearly dictated by the physical situation. For weak coupling, $U<J$ (not studied here, cf. \cite{hermanns_prb_14}), the second Born approximation is adequate. For moderate coupling, $U \le 2J$, proper approximations are the particle-particle $T$-matrix (TMA) and the third-order approximation (TOA), which was investigated here for the first time.
    \item For $U \le 2J$, the choice of $\Sigma$ depends on the local densities (filling): for densities close to zero (or close to one), TMA is appropriate, confirming earlier results for small clusters \cite{friesen09}, whereas near half filling TOA is significantly more accurate, as it contains contributions neglected in TMA.
    \item For the present system sizes the HF-GKBA (with the relevant selfenergy) yields more accurate results compared to the corresponding two-time simulations (due to the artificial damping observed in the latter). While the envelopes of global dynamical quantities (energies, cloud size, expansion velocity, density imbalance etc.) are captured very accurately, oscillations of these quantities are reproduced only qualitatively, for $U \le 2J$.
    \item Full two-time NEGF simulations can be used as a support of the HF-GKBA data, as typically the exact result is enclosed between the single-time and two-time simulations.
    One half of the difference of the two yields a (conservative) estimate of the numerical error, at least for couplings $U\lesssim 6 J$. 
%    (all considered setups) the correct description of oscillations is prevented by the artificial damping.

\end{enumerate}
%In general, the quality of the NEGF method depends on the choice of the selfenergy $\Sigma$. We showed that based on the system setup and the expected dynamics one can decide which choice of $\Sigma$ is best suited to provide reliable results. For the considered systems and coupling strengths the relevant approximations are given by the TMA (for small or large local densities) and the TOA (for dynamics near half-filling). 
%Furthermore, the best description of the dynamics is achieved by applying the HF-GKBA. This combined approach provides the correct envelope of dynamical quantities and properly describes oscillations at least for $U \le 2J$. In more detail, we found that for local quantities such as the local density or double occupation there might be a few slightly stronger deviations while for global quantities that describe the overall behavior such as the expansion velocity or the imbalance the relative error does not exceed \niclas{$\Delta=?$}. 
%
%Finally, our results confirm the high quality of the NEGF method from previous observations~\cite{} and generalize it to new setups and much larger system sizes. 
Based on this analysis of the NEGF capabilities, the main outcome of this paper is that NEGF and DMRG have, to a large degree, complementary strengths and limitations, with respect to the interaction strength. If $U$ does not exceed the bandwidth of the system, the NEGF approach has predictive power even for long-time propagations, and it is directly applicable to 2D and 3D systems~\cite{schluenzen16}. In contrast, the exponential spreading of entanglement narrows the DMRG approach to very short 1D simulations (somewhat larger times can be reached than presented here by using more states and possibly also by using variants of the algorithm \cite{enss2012,Haegeman2011}). On the other hand, If $U$ is larger than the bandwidth, the NEGF approach, in its present form, does not describe the dynamics properly, due to the built-in perturbative character of the approximations, whereas the DMRG method provides the exact dynamics for rather long times, although being limited to 1D and small 2D systems.

\subsection{Complementarity of NEGF and DMRG simulations: A case study}

\begin{figure}[tb]
\includegraphics[width=\columnwidth]{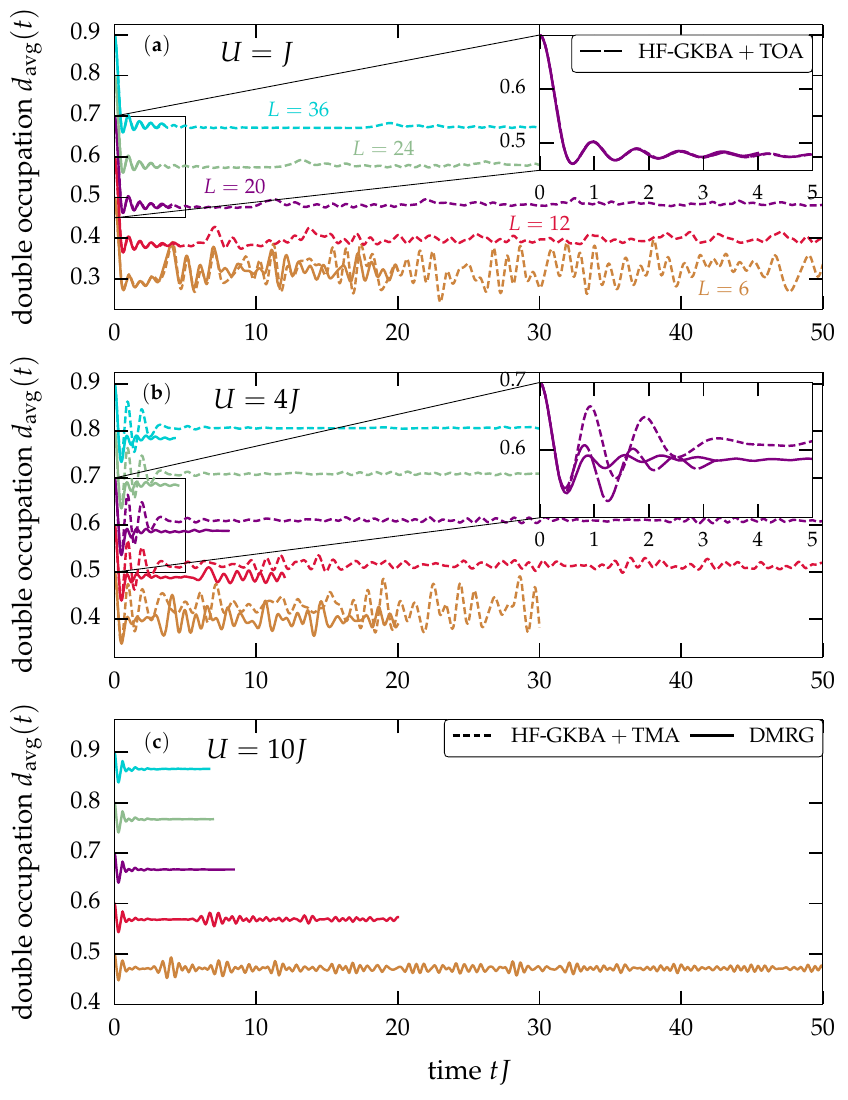}
\caption{(Color online)
{\it Relaxation of a CDW state of doublons.}
System-size dependence and long-time evolution of the average double occupancy, Eq.~(\ref{eq:d_avg}), for (a) $U=J$, (b) $U=4J$ and (c) $U=10J$ (DMRG results only) for chains of length $L=6,12,20,24,36$.  Full lines: DMRG, short dashes: HF-GKBA + TMA. The insets, in addition, show HF-GKBA+TOA results (long dashes). For better visibility, curves for different $L$ are shifted vertically by $0.1$.
}\label{fig:longtime}
\end{figure}

To illustrate this complementarity and  the reach of the two methods, we have performed additional long-time simulations and investigated the system size-dependence of the simulations, for the CDW setup (cf.~\sref{sec:cdw}). As a particularly sensitive quantity, we introduce the average double occupation 
\begin{equation}
{d_\tn{avg}(t) = L^{-1} \sum_{\bm{s}} d_{\sxbm} (t)}. 
\label{eq:d_avg}
\end{equation}
The time evolution of $d_\tn{avg}$ is shown in \fref{fig:longtime} for different chain lengths $L$, ranging from $6$ to $36$, corresponding to $6,\dots, 36$ particles, for (a) $U=J$, (b) $U=4J$ and (c) $U=10J$. The NEGF simulations use the HF-GKBA with $T$-matrix selfenergies. TOA simulations had a stability problem and are included only for shorter times (see insets).

%\footnote{We note, that in \fref{fig:longtime} instead of TOA the $T$-matrix selfenergy is chosen which appears to be more stable in the long-time limit.}. 
%
%
% To further demonstrate the accuracy and the reach of the considered methods, 
Starting from the case of $U=J$ [cf. \fref{fig:longtime}(a)], one can see that the short-time dynamics ($tJ<3$) of all considered chains are very similar. After the transient oscillations have decayed a quasi-stationary regime is observed. However, at some time the double occupation abruptly increases again [cf. \fref{fig:longtime}(a)]. These revivals occur periodically, becoming weaker with increasing system size $L$, and their periods increase nearly linearly with $L$. This indicates particles with a critical velocity that pass through the entire system. It should be noted that, after the revivals, $d_\tn{avg}$ starts to fluctuate inhomogeneously with an amplitude that decreases with $L$. We note that similar revivals and system-size dependencies were reported in Ref.~\onlinecite{Kiendl2016}.
% 
% By construction, the NEGF and DMRG method are complementary with respect to length of the time evolution depending on the interaction strength. 
While for $U=J$ the DMRG simulations are restricted to very short times, e.g., $tJ \sim 20$ for $L=6$ ($tJ\sim 5$ for $L=12$), the NEGF simulations easily allow to reach $tJ=50$ and more, for all chain lengths. The excellent agreement with DMRG is striking, suggesting that also the long time results are reliable. 

Consider now the case $U=4J$,~\fref{fig:longtime}(b). Here the complementary behavior of the two approaches becomes particularly obvious. While 
DMRG simulations show an improved performance compared to $U=J$ and reach times of the order of $tJ~\sim 20$, for $L=6$ ($tJ~\sim 12$, for $L=12$), NEGF simulations still reach times of the order of $tJ=50$, however, it is more difficult to achieve convergence. Simulations with the most accurate TOA selfenergy are stable only for short times, on the order of $tJ\sim 4$, for $L=20$, similar to DMRG, and are in good agreement with the latter [cf. inset of~\fref{fig:longtime}(b)]. Long-time simulations are presently possible only with $T$-matrix selfenergies which, however, exhibit a small upshift, compared to DMRG. 
Interestingly, the $L$-dependent revivals that were observed in the NEGF simulations for $U=J$ are confirmed here as well by the NEGF results and, even more clearly in the DMRG runs.

Finally, in \fref{fig:longtime}(c) we show results for $U=10J$. Here,  accurate long-time evolutions for small systems can be easily performed with the DMRG method. In contrast, the available NEGF approximations are not accurate enough and show poor convergence for long times (results from NEGF simulations not included in the figure).

\subsection{Outlook}

After this analysis of the NEGF approach and the illustration of the interesting complementarity with DMRG we briefly discuss questions that will be of interest for future developments.
First, it will be very important to extend the arsenal of selfenergies.
One important improvement will be achieved by extending the $T$-matrix approximation by including electron-hole contributions and by including dynamical screening effects (FLEX approximation~\cite{hermanns_phd}). These choices for the selfenergy will help to extend the interaction range where NEGF properly describes the dynamics. Another way to access larger $U$ is to derive novel selfenergies via a perturbation expansion with respect to $U^{-1}$, i.e., by starting from a Hamiltonian that includes doublons directly. Finally, it would be interesting to further improve the GKBA. While it was found to cure the artificial damping problems of two-time simulations, the dynamics is often too weakly damped. This behavior should improve if one uses instead of HF-propagators correlated propagators~\cite{kwong_pss_98,latini}.

While in this paper, 1D and small quasi-1D systems have been investigated for uncorrelated initial states, it will be interesting to extend the present method comparison to more complex, correlated initial states (including the ground states) as well as to larger 2D and 3D systems. It will also be interesting to analyze the dependence of the dynamics on the sign of the interaction~\cite{schneider12,schluenzen16}, to investigate disordered setups~\cite{Basko2006,Gornyi2005,Nandkishore2015,Altman2015} and to compare the fermionic simulations to those for bosonic lattice systems. Finally, the access of long simulation times by NEGF and DMRG for weak and strong coupling, respectively should allow one to study interesting features of the quantum-quench dynamics such as prethermalization
\cite{Berges2004,Moeckel2008,Eckstein2009}.

%%%%%%%%%%%%%%%%%%%%%%%%%%%%%%%%%%%%%%%%   Bibliography

%\bibliographystyle{apsrev}
%\bibliography{references}

\end{document}